\documentclass[%
onecolumn,
aps,%
showpacs,%
floatfix,%
superscriptaddress,%
nofootinbib%
]{revtex4}
\usepackage{graphicx,amsmath,amssymb}

\begin{document}


\preprint{hep/ph-0703053}

\title{Peak loops untying the degeneracy of the neutrino parameters}

\author{Masafumi~Koike}
\email{koike@krishna.th.phy.saitama-u.ac.jp}
\affiliation{%
  Physics Department, Saitama University,
  255 Shimo-Okubo, Sakura-ku, Saitama, Saitama 338-8570, Japan
}
\author{Masako~Saito}
\email{msaito@krishna.th.phy.saitama-u.ac.jp}
\affiliation{%
  Physics Department, Saitama University,
  255 Shimo-Okubo, Sakura-ku, Saitama, Saitama 338-8570, Japan
}

\date{\today}

\begin{abstract}
  Systematic analysis of the determination of the value of leptonic
  CP-violating angle $\delta$ and the neutrino mass hierarchy
  $\mathrm{sgn} \, \delta m^{2}_{31}$ by long baseline neutrino
  oscillation experiments is presented.
  We note the difficulty to distinguish a pair of oscillation
  probability spectra that are peaked at the same energy and have the
  same probability at that energy.
  We thereby set forth the peak-matching condition as a criterion of
  the presence of degeneracy, and visualize it by intersections of the
  trajectories drawn by a peak of an oscillation spectrum while the
  value of $\delta$ is varied from $0$ to $2\pi$.
  We numerically calculate the pairs of the trajectories for both
  hierarchies and show that the pair becomes disjoint as the baseline
  gets longer than a critical length, indicating the matter effect
  resolving the degeneracy on the hierarchy.
  We formulate the trajectories into analytic expressions and
  evaluate the critical length.
  We provide prospects of the following four approaches of resolving
  the hierarchy: making the baseline longer than the critical length,
  using both neutrinos and anti-neutrinos, combining experiments with
  different baseline lengths, and observing two or more oscillation
  peaks.
\end{abstract}

\maketitle

\section{Introduction}
The knowledge of the flavor structure of the lepton sector has been
vastly improved through the accumulation of experimental studies on
the neutrino oscillation \cite{bib:experiments}, yet it is still
incomplete; in the parametrization of reference \cite{Yao:2006px}, the
values of a mixing angle $\theta_{13}$, the mass hierarchy
$\mathrm{sgn} \, \delta m^{2}_{31}$, and the CP-violating angle
$\delta$ are still poorly known.
No lower bound of $\theta_{13}$ is given at present and the values of
$\mathrm{sgn} \, \delta m^{2}_{31}$ and $\delta$ are not known at all.
The leptonic CP-violation search is one of the most exciting
goal the neutrino experiments can offer.
It is done by observing the appearance channel such as $\nu_{\mu} \to
\nu_{\mathrm{e}}$, which is suppressed by a small factor of $\sin^{2}
2\theta_{13}$.
On this account, we expect the two-staged strategy in the search for
CP violation:
the first stage is the search for $\theta_{13}$ using nuclear
reactors \cite{bib:Nuclear-Planned} and
accelerators \cite{bib:Accelerator-Planned}, which currently targets
the sensitivity of $\sin^{2} 2\theta_{13} \sim O(10^{-2})$;
the second stage is the search for the sign of $\delta m^{2}_{31}$ and
for the CP violation, for which the next generations of long baseline
neutrino oscillation experiments will offer promising
opportunities \cite{bib:Accelerator-Planned}.

We focus upon the second stage, expecting the discovery of sufficiently
large value of $\theta_{13}$ in the first stage.
We consider the search for the CP-violating angle through long
baseline neutrino oscillation experiments using a conventional beam of
muon neutrinos.
The search for $\delta$ and $\mathrm{sgn} \, \delta m^{2}_{31}$ are
mutually entangled and the values of these two are not necessarily
determined uniquely by a single experiment with a fixed baseline
length, leaving a degeneracy of parameter values.
The parameter degeneracy obstructs the search for the parameter values
and should be avoided
\cite{%
  bib:degeneracy,%
  bib:two-baselines-combo,%
  bib:biprob-plot,%
  Barger:2001yr,%
  KoikeSaito%
}.
We introduce an intuitive illustration on the determination of the
parameters from the oscillation spectrum, and offer a view on the
emergence of degeneracy and its resolution.
We note that two oscillation spectra are difficult to distinguish and
likely to cause the degeneracy when they are peaked at the same energy
and the oscillation probabilities at the peak coincide.
We trace the oscillation peak varying the values of $\delta$ and
$\mathrm{sgn} \, \delta m^{2}_{31}$ to show how the search for them
is entangled and the degeneracy is invited.
We change the baseline length as well and illustrate from our point of
view how the degeneracy disappears when the baseline gets long.
We organize these analyses by deriving analytic expressions of the
oscillation probability at the peaks and offer an outlook of the
presence of degeneracy and possible ways to avoid it.

The outline of this paper is as follows.
In Sec.~\ref{sec:Peak-matching-cond}, we present the peak-matching
condition as a criterion for the presence of degeneracy and show an
organized view of the emergence of the degeneracy.
In Sec.~\ref{sec:Analytic-expressions}, we derive an analytic
expression for the $\nu_{\mu} \to \nu_{\mathrm{e}}$ appearance
probability at their peaks to show how the loops move and change the
shape in varying the baseline length.
In Sec.~\ref{sec:Solving-hierarchy-degeneracy}, we apply the peak
loops to the evaluation of four methods to avoid the degeneracy. 
Section \ref{sec:Conclusion-and-discussions} presents the conclusion
and discussions.

\section{Peak-matching condition and presence of degeneracy}
\label{sec:Peak-matching-cond}
\begin{figure}
\begin{center}
  \includegraphics[width=85mm]{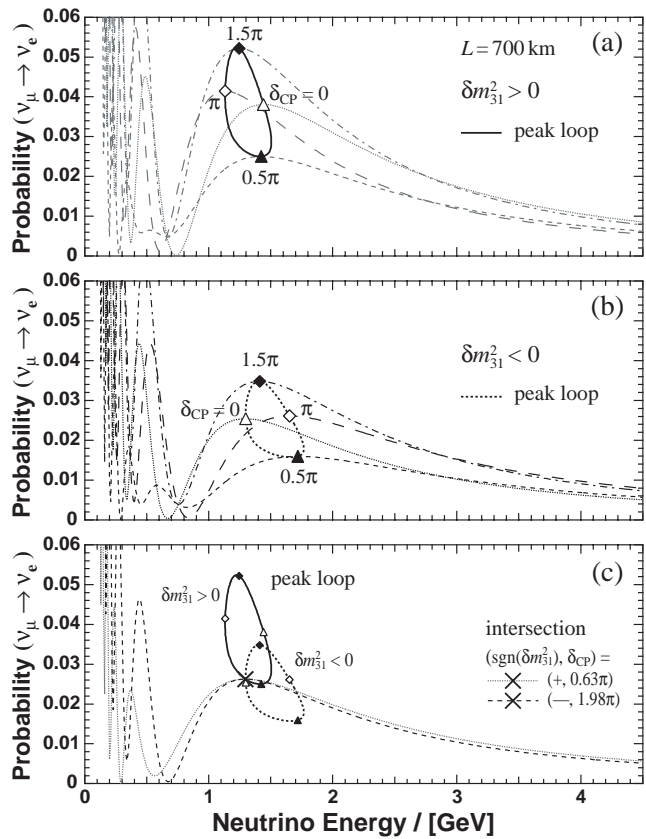}
  \caption{
    The $\nu_{\mu} \to \nu_{\mathrm{e}}$ appearance probabilities and
    the trajectories of their first peaks for the baseline length of
    $700 \, \mathrm{km}$.
    The oscillation parameters in Eq.~(\ref{eq:example-params}) are
    adopted.
    The top figure (a) is for the normal hierarchy and the middle
    (b) for the inverted, and each includes the probability
    spectra for $\delta = 0, \pi/2, \pi$ and $3\pi/2$.
    The bottom figure (c) reproduces the trajectories in (a) and (b)
    overlaid, along with two oscillation spectra peaked at the crossed
    intersection of these trajectories.
  }
  \label{fig:appearance-probs}
\end{center}
\end{figure}

We assume that the number of neutrino generations is three and use the
standard definitions of the quadratic mass differences $\delta
m^{2}_{ij}$ ($\{i, j\} \subset \{1, 2, 3\}$), the mixing angles
$\theta_{ij}$, and the CP-violating angle $\delta$ \cite{Yao:2006px}.
We consider a search for the value of $\delta$ and the mass hierarchy
$\mathrm{sgn} \, \delta m^{2}_{31}$ by observing $\nu_{\mu} \to
\nu_{\mathrm{e}}$ appearance probability in long baseline neutrino
oscillation experiments.
Let us illustrate with an example that $\nu_{\mu} \to
\nu_{\mathrm{e}}$ appearance probability enables us to pursue the sign
of $\delta m^{2}_{31}$ and the value of $\delta$.
We show in Fig.~\ref{fig:appearance-probs} the $\nu_{\mu} \to
\nu_{\textrm{e}}$ appearance probabilities for baseline length $L =
700 \, \mathrm{km}$.
The value of $\delta$ and the sign of $\delta m^{2}_{31}$ are varied
while other parameters are fixed to the following set of example
values allowed by the experimental limits \cite{Yao:2006px}:
\begin{subequations}
\begin{equation}
\begin{gathered}
  \delta m^{2}_{21} = 8.2 \times 10^{-5} \, \mathrm{eV}^{2} \, , \;
  \bigl\lvert \delta m^{2}_{31} \bigr\rvert
  = 2.5 \times 10^{-3} \, \mathrm{eV}^{2} \, , \;
  \\
  \sin^{2} 2\theta_{12} = 0.84 \, , \;
  \sin^{2} 2\theta_{23} = 1.0 \, , \;
  \sin^{2} 2\theta_{13} = 0.06 \, .
  \label{eq:example-params-dmm-and-theta}
\end{gathered}
\end{equation}
The matter density $\rho$ on the baseline is assumed to be constant
and take a reference value of
\begin{equation}
  \rho = 2.6 \, \mathrm{g/cm^{3}},
  \label{eq:example-params-rho}
\end{equation}
\label{eq:example-params}
\end{subequations}
which is related to the electron number density on the matter
$n_{\textrm{e}}$ as $n_{\textrm{e}} = N_{\textrm{A}} Y_{\textrm{e}}
\rho$ with the Avogadro constant $N_{\textrm{A}}$ and the
proton-to-nucleon ratio $Y_{\textrm{e}}$ on the baseline.
We take the values of Eq.~(\ref{eq:example-params}) with
$Y_{\textrm{e}} = 0.5$ in our following numerical calculations unless
otherwise noticed.
Figure \ref{fig:appearance-probs} shows the energy spectra of
appearance probability (a) for $\delta m^{2}_{31} > 0$ (normal
hierarchy) and (b) for $\delta m^{2}_{31} < 0$ (inverted hierarchy).
Figure \ref{fig:appearance-probs} (c) shows the probability for the
parameter values given in the figure.
We can clearly see in (a) and (b) the dependence of appearance
probability upon $\mathrm{sgn} \, \delta m^{2}_{31}$ and $\delta$
enables the search for these parameters through $\nu_{\textrm{e}}$
appearance experiments.

In spite of this dependence, experiments may allow parameter values in
more than one regions of parameter space and leave them with
degeneracy, which should be resolved
\cite{%
  bib:degeneracy,%
  bib:two-baselines-combo,%
  bib:biprob-plot,%
  Barger:2001yr,%
  KoikeSaito%
}.
The insight to the presence of degeneracy is enlightening for this
purpose, yet is difficult to gain due to complicated dependence of
experimental results on many oscillation parameters and controllable
setups.
Searching for a succinct criterion of its presence, we focus on the
peak of the oscillation probability spectrum.
The two spectra in Fig.~\ref{fig:appearance-probs} (c), having the
position of oscillation peaks in the $E$-$P$ plane in common at the
cross, show a notable similarity of the shapes over the energy range
above about $1 \, \mathrm{GeV}$, and are expected to be difficult to
distinguish by experiments with the typical visible energy $E > (0.5
\,\textrm{--}\, 1.0) \,\mathrm{GeV}$ of neutrinos.
This expectation can be confirmed by the $\chi^{2}$ goodness-of-fit
analyses \cite{Koike:2005dk}.
We put the above observation into the following statement of the
peak-matching condition as a criterion of the presence of the
degeneracy \cite{KoikeSaito}.
Let us denote the peaks of the appearance probability spectrum by $(
E_{n}, P_{n} )$ ($n = 0, 1, 2, \cdots$) with $E_{0} > E_{1} > E_{2} >
\cdots$, and suppose that the visible energy includes only one of
them.
The two sets of parameter values, collectively denoted by $\{
\vartheta_{i} \}$ and $\{ \vartheta'_{i} \}$, are expected to be
degenerate degenerate when the peak-matching condition
\begin{subequations}
\begin{align}
  E_{n} \bigl( \{ \vartheta_{i} \} \bigr)
  & =  E_{n} \bigl( \{ \vartheta'_{i} \} \bigr) \, ,
  \label{eq:epeak-matching-cond}
  \\
  P_{n} \bigl( \{ \vartheta_{i} \} \bigr)
  & =  P_{n} \bigl( \{ \vartheta'_{i} \} \bigr)
  \label{eq:ppeak-matching-cond}
\end{align}
\label{eq:peak-matching-cond}
\end{subequations}
is satisfied by this visible peak.

We visualize this condition by introducing a trajectory made by an
oscillation peak as the value of $\delta$ varied from $0$ to $2 \pi$
while other parameters kept fixed.
The position of a peak for the normal (inverted) hierarchy moves
clockwise (counterclockwise) as the value of $\delta$ increases.
A trajectory of a peak forms a closed loop as presented in
Fig.~\ref{fig:appearance-probs} (a) and (b) for the first peak ($n =
0$) of respective hierarchies, and intersects with the loop for the
other hierarchy as seen in Fig.~\ref{fig:appearance-probs} (c), where
one of them are marked by a cross.
The parameter values corresponding to the intersections are the ones
that satisfy the peak-matching condition.
In the example of Fig.~\ref{fig:appearance-probs} (c), the pairs of
the values of $(\mathrm{sgn} \, \delta m^{2}_{31}, \delta)$ that give
the intersections are
\begin{equation}
  \bigl\{ (+, 0.63\pi) \, , (-, 1.98 \pi) \bigr\} \; , \quad
  \bigl\{ (+, 0.12\pi) \, , (-, 1.34 \pi) \bigr\} \, .
  \label{eq:ex-degenerate-pair}
\end{equation}
If the true set of parameter values is one of the above along with
Eq.~(\ref{eq:example-params-dmm-and-theta}), the experiments would
leave the degeneracy with the other of the peak-matching pair.
The parameter values corresponding to the paralleling sides of the
loops would be also degenerate, especially when these sides are
indistinguishable due to the limited experimental resolution.

\begin{figure}
\begin{center}
  \includegraphics[width=85mm]{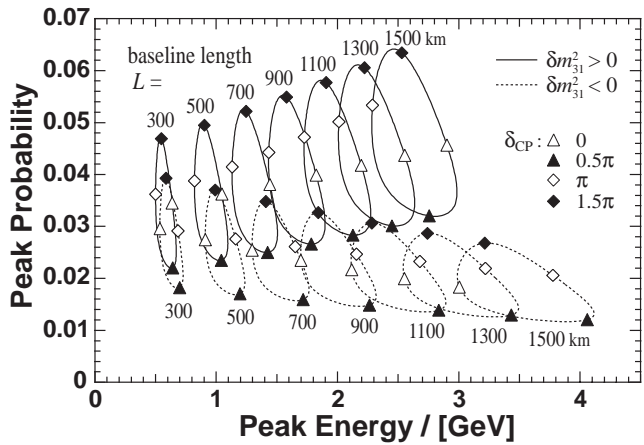}
  \caption{
    Trajectories of the first oscillation peak for the baseline lengths
    between $300 \, \mathrm{km}$ and $1500 \, \mathrm{km}$.
    The oscillation parameters in Eq.~(\ref{eq:example-params}) are
    adopted.
    The solid line is for the normal hierarchy and the dotted one
    for the inverted.
  }
  \label{fig:nu_trajectory}
\end{center}
\end{figure}

Equipped by the peak loops, we apply them to examine how the
degeneracy emerges and disappears as we vary the baseline length of
the neutrino oscillation experiments.
We present in Fig.~\ref{fig:nu_trajectory} peak loops for baseline
length from $300 \, \mathrm{km}$ to $1500 \, \mathrm{km}$.
For a relatively short baseline ($L \sim 300 \mathrm{km}$ for our
example),
the loop for the normal hierarchy (``normal loop'') and 
         for the inverted hierarchy (``inverted loop'')
have similar shape and lie significantly overlapped with each other.
As a baseline gets longer, the normal loop and the inverted loop move
right-upward and left-downward, respectively, and diverge due to the
matter effect.
The normal loop is expanded in the $E$- and the $P$-direction, while
the inverted loop is appreciably stretched in the $E$-direction and
flattened in the $P$-direction.
The pair of loops becomes disjoint at a critical length
$L_{\textrm{crit}}$, which is in between $1100 \,\textrm{--}\, 1300 \,
\mathrm{km}$ in this example as seen in Fig.~\ref{fig:nu_trajectory}.

The study so far leads us to the following outlooks upon determining
the CP-violating angle and the mass hierarchy.
In a short baseline length, the pair of loops overlaps considerably
and consequently the hierarchy is difficult to determine by
experiments.
The search for $\delta$ in this case is entangled with that for the
hierarchy, and its value is also difficult to determine.
Another obstacle is invited by the limitation on the energy resolution
which may not be enough to distinguish the paralleling sides of the
narrow loops.
As the baseline gets longer, the pair of loops becomes separated and
the determination of the hierarchy is made easier.
The search for the CP-violating angle becomes disentangled from the
hierarchy search at the same time.
The flattened shape of the inverted loops for a long baseline gives
rise to the requirement on precise and accurate measurements of the
oscillation probability to avoid introducing an extra degeneracy.

\section{Analytic study of the peak loops}
\label{sec:Analytic-expressions}
In this section, we derive an analytic expression of the oscillation
probability to formulate properties of the peak
loops \cite{KoikeSaito}.
We thereby show how a pair of peak loops for $\delta m^{2}_{31}
\gtrless 0$ distort and are separated as the baseline length
increases.

We derive a formula of the oscillation probability by treating
$\Delta_{21} \equiv \delta m^{2}_{21} L / 2E$ and $\Delta_{\textrm{m}}
\equiv \sqrt{2} G_{\textrm{F}} n_{\textrm{e}} L = \sqrt{2}
G_{\textrm{F}} N_{\textrm{A}} Y_{\textrm{e}} \rho L$ as perturbation
parameters, where $G_{\textrm{F}}$ is the Fermi constant, and expand
the $S$-matrix systematically \cite{bib:AKS-related}.
We calculate the $\nu_{\mu} \to \nu_{\textrm{e}}$ appearance
probability to the second order to investigate the distortion of the
peak loops in varying the baseline length.
The result turns out to be lengthy, and we apply an additional
approximation considering the smallness of $\theta_{13}$ and the
relation $\Delta_{\textrm{m}} > \Delta_{21}$, which holds for the
cases we are interested in: we drop $O(\sin^{2} \theta_{13})$-terms in
the coefficients of $\Delta_{21}^{2}$ and $\Delta_{\textrm{m}}
\Delta_{21}$ as well as $O(\sin^{3} \theta_{13})$-terms in that of
$\Delta_{\textrm{m}}^{2}$.
We then obtain
\begin{subequations}
\begin{equation}
  P(\nu_{\mu} \rightarrow \nu_{\mathrm{e}}, E) 
  = 4l (A \sin^{2} \Theta + B) \, ,
\label{eq:Pm2e-aks-order2-body}
\end{equation}
where
\begin{eqnarray}
  A
  & = &
    1 + 2\frac{\Delta_{\textrm{m}}}{\Delta_{31}} (1-2s_{13}^2) 
  - \Delta_{21}\frac{j}{l}\sin\delta
  + 3 \frac{\Delta_{\textrm{m}}^{2}}{\Delta_{31}^{2}}
  \nonumber \\ & &
  - \Delta_{21} \frac{\Delta_{\textrm{m}}}{\Delta_{31}} \frac{j}{l}
    \Bigl( \sin\delta + \frac{\Delta_{31}}{2} \cos\delta \Bigr)
  + \frac{\Delta_{21}^2}{2} \frac{j}{l}
    \biggl[ \frac{j}{l}\cos\delta +(1 - 2s_{12}^{2}) \biggr] \cos\delta  \, ,
\label{eq:Pm2e-aks-order2-A}
\end{eqnarray}
\begin{eqnarray}
  \Theta
  & = &
    \frac{\Delta_{31}}{2} 
  - \frac{\Delta_{\textrm{m}}}{2}(1 - 2s_{13}^{2} )
  + \frac{\Delta_{21}}{2} \Bigl( \frac{j}{l}\cos\delta - s_{12}^{2} \Bigr)
  \nonumber \\ & &
  - \frac{\Delta_{21}}{2} \frac{\Delta_{\textrm{m}}}{\Delta_{31}} \frac{j}{l} 
    \Bigl( \cos\delta + \frac{\Delta_{31}}{2} \sin\delta \Bigr)
  + \frac{\Delta_{21}^2}{2} \frac{j}{l}
    \biggl[ \frac{j}{l}\cos\delta + \frac{1}{2}(1 - 2s_{12}^{2}) \biggr]
    \sin\delta \, ,
\label{eq:Pm2e-aks-order2-Theta}
\end{eqnarray}
and
\begin{equation}
  B = \frac{\Delta_{21}^{2}}{4} \frac{j^{2}}{l^{2}} \sin^{2}\delta \, ,
\label{eq:Pm2e-aks-order2-B}
\end{equation}
\label{eq:Pm2e-aks-order2}
\end{subequations}
with
$l = c_{13}^{2} s_{13}^{2} s_{23}^{2}$,
$j = c_{13}^{2} s_{13} c_{23} s_{23} c_{12} s_{12}$,
$\Delta_{ij} = \delta m^{2}_{ij} L/2E$,
$s_{ij} = \sin\theta_{ij}$, and $c_{ij} = \cos\theta_{ij}$.
The approximation employed is suitable especially for a short
baseline, typically for $L < O(10^{3} \, \mathrm{km})$, and is valid
at least marginally for our purpose.

We calculate the maximum of the oscillation probability of
Eq.~(\ref{eq:Pm2e-aks-order2}) and obtain the energy at the peak as
\begin{eqnarray}
  E_{n}
   =
  \frac{|\delta m^{2}_{31}| L}{2\Pi}
  &\Biggl\{&
    \biggl[
          1
      \mp \Delta_{\textrm{m}}
          \frac{1}{\Pi} \Bigl( 1 - \frac{4}{\Pi^{2}} \Bigr)
          \bigl(1 - 2 s_{13}^{2} \bigr)
      \mp R s_{12}^{2}
      \nonumber \\
      &&\hspace*{1cm}
      + \Delta_{\textrm{m}}^{2} \frac{1}{\Pi^{2}}
        \Bigl( 1 - \frac{12}{\Pi^{2}} + \frac{48}{\Pi^{4}} \Bigr)
      - R^{2} \frac{1}{2} \Bigl( 1 - \frac{4}{\Pi^{2}} \Bigr)
        \frac{j^{2}}{l^{2}}
    \biggr]
    \nonumber \\
    &+& R
    \biggl[
      \pm 1
        - \Delta_{\textrm{m}} \frac{1}{\Pi} \Bigl( 1 - \frac{8}{\Pi^{2}} \Bigr)
        - 2 R (1 - 2 s_{12}^{2})
    \biggr]
    \frac{j}{l}
    \cos \delta
    \nonumber \\
    &+& R \frac{2}{\Pi}
    \biggl[
          1
      \mp \Delta_{\textrm{m}} \frac{\Pi}{4}
          \Bigl( 1 + \frac{8}{\Pi^{2}} - \frac{64}{\Pi^{4}} \Bigr)
      \pm R \frac{\Pi^{2}}{4} (1 - 2 s_{12}^{2})
    \biggr]
    \frac{j}{l}
    \sin \delta
    \nonumber \\
    &-& R^{2}
        \frac{3}{2} \Bigl( 1 + \frac{4}{3} \frac{1}{\Pi^{2}} \Bigr)
        \frac{j^{2}}{l^{2}}
        \cos 2 \delta
    \pm R^{2} \frac{\Pi}{2} \frac{j^{2}}{l^{2}} \sin 2 \delta
  \Biggr\} \, ,
\label{eq:PeakE_2nd}
\end{eqnarray}
where $n = 0, 1, 2, \cdots$ is the peak index, $\Pi \equiv (2n + 1)\pi$, $R
\equiv \delta m^{2}_{21}/|\delta m^{2}_{31}|$, and the top of the double sign
is for $\delta m^{2}_{31} > 0$ and the bottom for $\delta m^{2}_{31} <
0$.
The oscillation probability at the peak is then given by
\begin{eqnarray}
  P_{n}
  =
  4l
  &\Biggl\{&
    \biggl[
          1 
      \pm \Delta_{\textrm{m}} \frac{2}{\Pi} \bigl( 1 - 2 s_{13}^{2} \bigr)
        + R^{2} \frac{3}{8} \Pi^{2}
          \Bigl( 1 + \frac{4}{3} \frac{1}{\Pi^{2}} \Bigr)
          \frac{j^{2}}{l^{2}}
        + \Delta_{\textrm{m}}^{2} \frac{1}{\Pi^{2}}
          \Bigl( 1 + \frac{4}{\Pi^{2}} \Bigr)
    \biggr]
    \nonumber \\
   &-&
    R \frac{\Pi^{2}}{2}
    \biggl[
        \Delta_{\textrm{m}} \frac{1}{\Pi} \Bigl( 1 - \frac{4}{\Pi^{2}} \Bigr)
      - R (1 - 2 s_{12}^{2})
    \biggr]
    \frac{j}{l} \cos \delta
    \nonumber \\
   &-&
    R \Pi
    \biggl[
          1
      \pm \Delta_{\textrm{m}} \frac{2}{\Pi} \Bigl( 1 - \frac{2}{\Pi^{2}} \Bigr)
      \pm R s_{12}^{2}
    \biggr]
    \frac{j}{l} \sin \delta
    \nonumber \\
    &+&
    R^{2} \frac{\Pi^{2}}{8} \Bigl( 1 - \frac{4}{\Pi^{2}} \Bigr)
    \frac{j^{2}}{l^{2}}
    \cos 2 \delta
    \pm
    R^{2} \frac{\Pi}{2} \frac{j^{2}}{l^{2}} \sin 2 \delta
  \Biggr\} \, .
\label{eq:PeakP_2nd}
\end{eqnarray}
The peak loops for $n = 0$ is obtained by keeping track of $(E_{n},
P_{n})$ as the value of $\delta$ is varied over a cycle from $0$ to
$2\pi$ while other parameters are fixed.

Taking another step ahead, we use these expressions to formulate
properties of the peak loop and their dependence on the oscillation
parameters and the baseline length.
We define the central position of the peak loop by the average values
of $E_{n}$ and $P_{n}$ over a cycle of $\delta$.
It is obtained by picking the terms independent of $\delta$ in
Eqs.~(\ref{eq:PeakE_2nd}) and (\ref{eq:PeakP_2nd}), and is given up to
the first order by
\begin{eqnarray}
  &&
  \bigl( \overline{E}_{n}, \overline{P}_{n} \bigr)
  =
  \nonumber \\
  &&
  \biggl(
    \frac{|\delta m^{2}_{31}| L}{2\Pi}
    \Bigl[
          1
      \mp \frac{\Delta_{\textrm{m}}}{\Pi}
          \Bigl( 1 - \frac{4}{\Pi^{2}} \Bigr) (1 - 2s_{13}^{2})
      \mp  R s_{12}^{2}
    \Bigr],
    4l
    \Bigl[
      1 \pm \frac{2 \Delta_{\textrm{m}}}{\Pi} (1 - 2s_{13}^{2})
    \Bigr]
  \biggr),
\label{eq:loop_center_nu}
\end{eqnarray}
where the double sign should be understood as in
Eqs.~(\ref{eq:PeakE_2nd}) and (\ref{eq:PeakP_2nd}).
The overall factor of $\overline{E}_{n}$ is proportional to the
baseline length and consequently drives the loop to the rightward in
the $E$-$P$ plane as the baseline gets longer.
The subleading matter-effect term is also proportional to the baseline
length, shifting the normal loop left-upward and the inverted one
right-downward.
These features are indeed seen in Fig.~\ref{fig:nu_trajectory}.

The width $\Delta E_{n}$ and the height $\Delta P_{n}$ of a loop are
estimated by taking the difference of the maximum and minimum values
of $E_{n}$ and $P_{n}$, and are obtained up to the second order as
\begin{subequations}
\begin{align}
  &
  \Delta E_{n}
  =
  |\delta m^{2}_{31}| L R \frac{1}{\Pi} \sqrt{ 1 + \frac{4}{\Pi^{2}} } \frac{j}{l}
  \biggl[ 1 \mp \Delta_{\textrm{m}} \frac{2}{\Pi}
                \frac{ 1 - 32/\Pi^{4} }{ 1 + 4/\Pi^{2} }
            \mp R \frac{1 - 2s_{12}^{2}}{1 + 4/\Pi^{2}}
  \biggr] \, ,
  \label{eq:Delta_E_nu}
  \\
  &
  \Delta P_{n}
  =
  8R\Pi j
  \biggl[
    1 \pm \Delta_{\textrm{m}} \frac{2}{\Pi} \Bigl( 1 - \frac{2}{\Pi^{2}} \Bigr)
      \pm Rs_{12}^{2}
  \biggr] \, .
  \label{eq:Delta_P_nu}
\end{align}
\label{eq:Delta_EP_nu}
\end{subequations}
The overall factor of $\Delta E_{n}$ widens the loop due to its
proportionality to the baseline length, and the matter effect offsets
the widening for the normal hierarchy and enhances it for the
inverted.
The dependence of $\Delta P_{n}$ is supplied by the matter-effect
term, which heightens the normal loop and flattens the inverted one.
These behaviors of the loops are also observed in
Fig.~\ref{fig:nu_trajectory}.

\begin{figure}
\begin{center}
  \includegraphics[width=85mm]{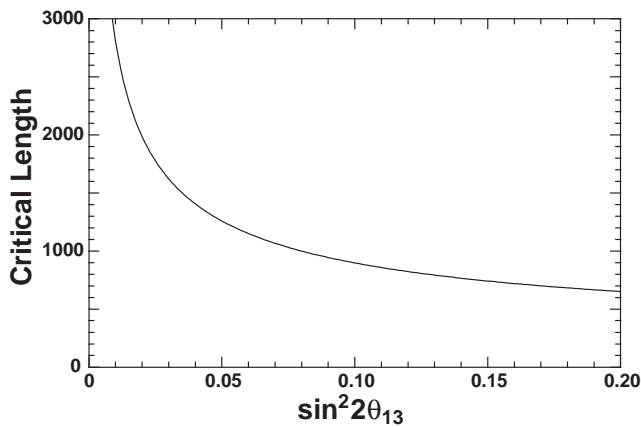}
  \caption{
    The dependence of the critical length on $\sin^{2} 2\theta_{13}$
    calculated by the first-order approximation formula of
    Eq.~(\ref{eq:Lcrit-def}).
    The oscillation parameters in Eq.~(\ref{eq:example-params}) are
    adopted except for the value of $\theta_{13}$.
  }
  \label{fig:critical-length}
\end{center}
\end{figure}

The critical baseline length can also be analyzed by the peak loops.
It is defined as the maximum length that keeps the intersection of the
normal and the inverted loops, and is calculated in the first-order
approximation as
\begin{eqnarray}
  L_{\textrm{crit}}
  &=&
  \frac{1}{\sqrt{2} G_{\textrm{F}} n_{\textrm{e}}} R
  \frac{\Pi}{ 1 - 12\Pi^{-2} + 64\Pi^{-4} }
  \frac{c_{23} c_{12} s_{12}}{s_{13} s_{23}} 
  \frac{1}{1 - 2 s_{13}^{2}}
  \nonumber \\ && \times 
  \Biggl[
    \sqrt{
      1 - \dfrac{12}{\Pi^{2}} + \dfrac{64}{\Pi^{4}}
        - \frac{4}{\Pi^{2}} 
          \Bigl( \frac{s_{13} s_{23} s_{12}}{c_{23} c_{12}} \Bigr)^{2}
    }
    -  
    \Bigl( 1 - \dfrac{8}{\Pi^{2}} \Bigr)
    \frac{s_{13} s_{23} s_{12}}{c_{23} c_{12}}
  \Biggr].
\label{eq:Lcrit-def}
\end{eqnarray}
The critical length is inversely proportional to $\sin \theta_{13}$
owing to the factor $c_{23} c_{12} s_{12}/s_{13} s_{23}$, with the
small correction of $O(\theta_{13})$.
This length can be indefinitely long since the lower bound on the
value of $\sin \theta_{13}$ is still unknown.
Figure \ref{fig:critical-length} shows the critical length as a
function of $\sin^{2} 2\theta_{13}$ for the first peak, where other
parameters are fixed to the example values of
Eq.~(\ref{eq:example-params}).
This graph reads $L_{\textrm{crit}} = 1150 \, \mathrm{km}$ for
$\sin^{2} 2\theta_{13} = 0.06$.
These results are qualitatively consistent with the result presented
in the reference \cite{Barger:2001yr}.

\section{Prospects of resolving the degeneracy through the peak loops}
\label{sec:Solving-hierarchy-degeneracy}
We consider possible methods to escape from the hierarchy degeneracy
in the long baseline experiments \cite{KoikeSaito}.

An experiment with baseline length longer than the critical length is
a promising for this purpose, being free from the hierarchy
degeneracy.
Nonetheless the very long baseline also brings on the challenge such
as small flux of the neutrino beam and possible large ambiguity of
matter effects.

Experiments with shorter baseline are more feasible, but leaves the
possibility of degeneracy.
Combining the observations of two or more peaks in the search avails
against this problem.
We consider the following three approaches and examine them in order%
\footnote{
  Enumeration starts from (2), reserving (1) for the approach
  employing the baseline longer than the critical length described
  above.
}:
(2) observing $\bar{\nu}_{\mu} \to \bar{\nu}_{\mathrm{e}}$ appearance
events in addition to $\nu_{\mu} \to \nu_{\mathrm{e}}$ events;
(3) doing two or more experiments with different baseline lengths;
and 
(4) doing an experiment that can observe two or more oscillation
peaks.

\begin{figure}
\begin{center}
  \includegraphics[width=85mm]{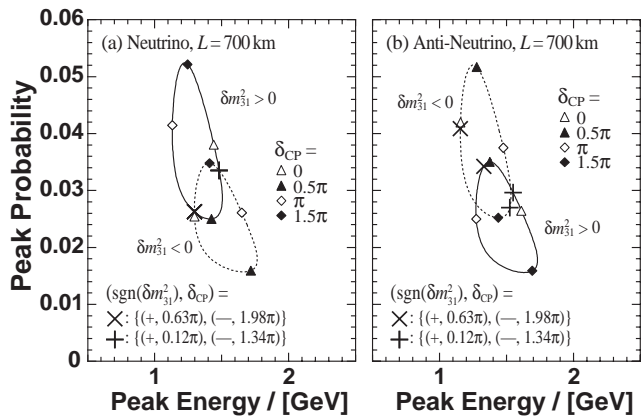}
  \caption{
    Trajectories of the first peak of (a) $\nu_{\mu} \to
    \nu_{\textrm{e}}$ probability and (b) $\bar{\nu}_{\mu} \to
    \bar{\nu}_{\textrm{e}}$ probability for the baseline length of
    $700 \, \mathrm{km}$.
    The solid and dotted lines are for the normal and inverted
    hierarchy, respectively.
    The cross and the plus sign in (a) are on the intersections of
    the  peak loops, and the corresponding values of parameters are
    also shown.
    The signs in (b) mark the points corresponding to these
    parameter values.
  }
  \label{fig:peak_loops_nu_nubar}
\end{center}
\end{figure}

The second approach employs both neutrinos and anti-neutrinos.
We present in Fig.~\ref{fig:peak_loops_nu_nubar} two pairs of peak
loops, one (a) for neutrinos and the other (b) for anti-neutrinos,
with the parameter values of Eq.~(\ref{eq:example-params}) and $L =
700 \, \mathrm{km}$.
Marked by a cross and a plus sign in
Fig.~\ref{fig:peak_loops_nu_nubar} (a) are the two intersections of
the normal and inverted loops.
Each intersection corresponds to the hierarchy degeneracy through two
sets of $(\mathrm{sgn} \, \delta m^{2}_{31}, \delta)$.
Turning to the loop for anti-neutrinos, these sets of parameters no
longer correspond to intersections but to two distinct points on the
peak loops of anti-neutrinos, as shown in
Fig.~\ref{fig:peak_loops_nu_nubar} (b) with the corresponding marks.
The combined analysis of $\nu_{\mu} \to \nu_{\mathrm{e}}$ events and
$\bar{\nu}_{\mu} \to \bar{\nu}_{\mathrm{e}}$ events therefore helps
resolve the hierarchy degeneracy.
The two plus signs for anti-neutrinos are close to each other on the
$E$-$P$ plane in our present example, and the resolution needs to be
high enough to distinguish the two points in order to make full use of
the anti-neutrino channel in resolving the degeneracy.

\begin{figure}
\begin{center}
  \includegraphics[width=85mm]{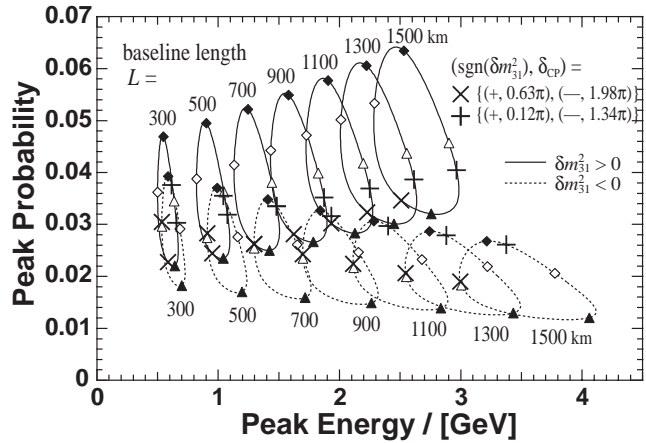}
  \caption{
    The points correspond to the two sets of parameter values shown
    in Fig.~\ref{fig:peak_loops_nu_nubar}, plotted on top of a
    series of loops reproduced from Fig.~\ref{fig:nu_trajectory}.
    Crosses and plus signs correspond to two intersections of the
    loops for $L = 700 \, \mathrm{km}$.
    The oscillation parameters in Eq.~(\ref{eq:example-params}) are
    adopted.
    The solid (dotted) line is for the normal (inverted) hierarchy.
  }
  \label{fig:loops-various-L}
\end{center}
\end{figure}

The third approach makes use of two or more different baseline
lengths \cite{bib:two-baselines-combo}.
We reproduce in Fig.~\ref{fig:loops-various-L} a series of peak loops
shown in Fig.~\ref{fig:nu_trajectory}.
On top of them, we mark the points correspond to the parameter values
for the intersections of the loops for $L = 700 \, \mathrm{km}$ as we
did in Fig.~\ref{fig:peak_loops_nu_nubar}.
We see that the crosses and the plus signs flows away from the
intersections as the baseline length deviates away from $700 \,
\mathrm{km}$.
The degeneracy can be thus resolved by an additional experiment with
two different baseline length, where the difference of the two
baseline lengths should be preferably large.

The fourth approach makes use of the second peak of appearance
probability in addition to the first one \cite{Diwan:2003bp}.
The energy of the second peak is approximately only one third of that
of the first peak, and we need sufficiently long baseline to make the
second peak observable.
\begin{figure}
\begin{center}
  \includegraphics[width=85mm]{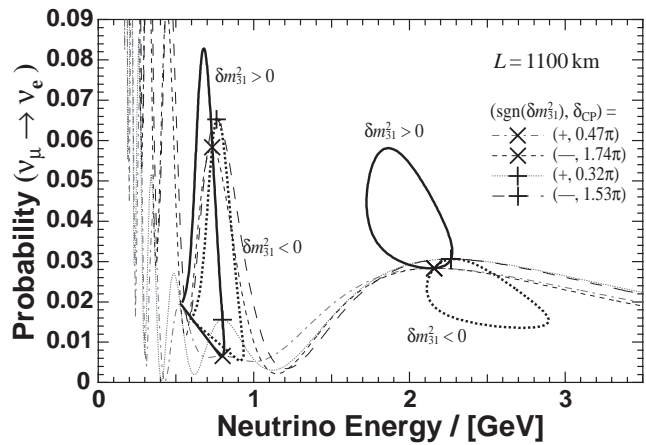}
  \caption{
    Trajectories of the first and the second oscillation peak for $L
    = 1100 \, \mathrm{km}$.
    The oscillation parameters in Eq.~(\ref{eq:example-params}) are
    adopted.
    The solid (dotted) loop is for the normal (inverted) hierarchy.
    Crosses and plus signs correspond to the sets of parameter
    values that bring the first peak at the intersection of its
    trajectories.
    The points for these values are also on the second
    peak loops and marked by the corresponding sign.
    Oscillation spectra for the parameter values are overlaid.
  }
  \label{fig:2ndpeak_nu_trajectory}
\end{center}
\end{figure}
We show pairs of loops for the first and the second peaks in
Fig.~\ref{fig:2ndpeak_nu_trajectory} for $L = 1100 \, \mathrm{km}$,
where we find that the loops for the second peak are more stretched in
the $P$-direction and overlap more significantly than the first-peak
loops owing to the factor $\Pi \equiv (2n + 1)\pi$ appearing in
Eqs.~(\ref{eq:loop_center_nu}) and (\ref{eq:Delta_EP_nu}).
Crosses and plus signs correspond to values at the two intersections
of the loops for the first peak.
The points of same sign are separated from each other on the second
peak loops and the hierarchy degeneracy will be thus removed by using
two peaks.

\section{Conclusion and discussions}
\label{sec:Conclusion-and-discussions}
We studied the search for the leptonic CP-violating angle $\delta$ and
the neutrino mass hierarchy $\mathrm{sgn} \, \delta m^{2}_{31}$ by an
observation of the $\nu_{\mu} \to \nu_{\textrm{e}}$ oscillation with
long baseline experiments.
The energy spectrum of $\nu_{\mu} \to \nu_{\textrm{e}}$ appearance
probability gives clues to the values of these parameters, but may
lead to degeneracy when the two spectra corresponding to the two
different parameter values are indistinguishable.
We implement the presence of degeneracy using the following criterion:
the oscillation probabilities for the two parameter values are peaked
at the same energy and have the same peak probability.
From the viewpoint of this peak-matching condition, we examined
systematically the presence of the degeneracy varying the value of
$\delta$, the sign of $\delta m^{2}_{31}$, and the baseline length.
We have also studied how the degeneracy is resolved by a single
experiment and a combination of experiments.

We have shown that the loops traced by the peak position offer
intuitive view on the parameter degeneracy.
A pair of disjoint loops indicate the absence of degeneracy while
intersecting loops do the presence, and each of intersections
corresponds to two sets of parameter values that are degenerate.
Deriving analytic expressions of the loop, we systematically studied
their dependence on the mass parameters, the mixing parameters, and
the baseline length.
We have seen that a pair of loops with different hierarchies are
completely separated when the baseline is longer than the critical
length, which is typically about $1000 \, \mathrm{km}$ or longer
and inversely proportional to $\sin \theta_{13}$.
We cited four approaches for resolving the degeneracy by long baseline
experiments:
taking the baseline longer than the critical length, %
employing anti-neutrinos as well as neutrinos, %
using two or more different baseline lengths, %
and carrying out experiments covering more than one oscillation peaks.
We went into each idea in terms of the peak loops and mentioned that
the second approach using anti-neutrinos may require a high resolution
to take advantage of using both neutrinos and anti-neutrinos.

The peak loops we employed have an equal significance to the
trajectories in the bi-channel plots introduced in the
reference \cite{bib:biprob-plot}, but simultaneously have differences
from them:
our plot employs only a single channel and thus can be applied to
single-channel experiments to explore possibilities their
possibilities;
we make essential use of the oscillation peak as a representative of
the spectrum over a finite energy range, while the bi-channel plot is
drawn for an arbitrary fixed energy or for integrated values over an
energy range at one's convenience.

We have kept the values of parameters fixed except for $\delta$ and
$\mathrm{sgn} \, \delta m^{2}_{31}$, assuming that these values would
be settled in advance of experiments.
Their ambiguities may, however, persist at the time of experiments and
hinder the searches
\cite{%
  bib:degeneracy,%
  bib:two-baselines-combo,%
  bib:biprob-plot,%
  Barger:2001yr,%
  Koike:2005dk%
}.
We can apply the peak loop to a comprehensive understanding of effects
of these ambiguities also.  We leave them for future works.

\section*{Acknowledgements}

The authors are grateful to Professor Joe Sato for his encouragement
during this work.


%
\end{document}